# Stress transfer mechanisms at the sub-micron level for graphene/ polymer systems

*George Anagnostopoulos[1], Charalampos Androulidakis[1,2], Emmanuel N. Koukaras[1], Georgia Tsoukleri[1], Ioannis Polyzos[1], John Parthenios[1], Konstantinos Papagelis[1,2]\* and Costas Galiotis[1,3]\**

[1]Institute of Chemical Engineering Sciences, Foundation for Research and Technology – Hellas (FORTH/ ICE-HT), P.O. Box 1414, Patras 265 04, Greece

[2]Department of Materials Science, University of Patras, Patras 26504, Greece

[3]Department of Chemical Engineering, University of Patras, Patras 26504, Greece

**ABSTRACT**

The stress transfer mechanism from a polymer substrate to a nano-inclusion, such as a graphene flake, is of extreme interest for the production of effective nanocomposites. Previous work conducted mainly at the micron scale has shown that the intrinsic mechanism of stress transfer is shear at the interface. However, since the interfacial shear takes its maximum value at the very edge of the nano-inclusion it is of extreme interest to assess the effect of edge integrity upon axial stress transfer at the submicron scale. Here, we conduct a detailed Raman line mapping near the edges of a monolayer graphene flake which is simply supported onto an epoxy based photoresist (SU8)/poly(methyl methacrylate) (PMMA) matrix at steps as small as 100 nm. We show for the first time that, the distribution of axial strain (stress) along the flake deviates somewhat from the classical shear-lag prediction for a region of about 2 μm from the edge. This behavior is mainly attributed to the presence of residual stresses, unintentional doping and/or edge effects (deviation from the equilibrium values of bond lengths and angles, as well as different edge chiralities). By considering a simple balance of shear-to-normal stresses at the interface we are able to directly convert the strain (stress) gradient to values of interfacial shear stress for all the applied tensile levels without assuming classical shear-lag behavior. For large flakes a maximum value of interfacial shear stress (ISS) of 0.4 MPa is obtained prior to flake slipping

# Introduction

Exfoliated graphene (EG) has been proven to be a perfect 2D crystal exhibiting unparalleled properties in a number of fields[1]. In spite of its irregular shape, EG is considered as a reference material against which graphenes produced by other methods, such as, Chemical Vapor Deposition (CVD)[2, 3], are compared with. Regarding the latter, it is widely acknowledged that CVD graphene is of inferior quality than exfoliated flakes[4, 5]. Samples produced by CVD do suffer from intrinsic defects such as folds, holes etc. Furthermore, the formation of grain boundaries[6] of micrometer dimensions in CVD graphene membranes affects adversely their mechanical integrity and performance [4]. On the contrary, EG quality still remains much more appealing[1], not only as a model material for theoretical studies but also as a model single grain system[7]. New exfoliation procedures can yield flakes over 100 µm in dimensions[8-10]. However, EG production is slow and labour intensive and therefore is unsuitable for automated production.

For EG exfoliation which is done manually, the involvement of an operator makes the process rather difficult to control and, therefore, quite often the electrical and mechanical response of the material is affected by the quality and the degree of doping (through charged impurities) of the flake, as well as, the presence of residual mechanical fields. One area for which graphene shows promise is its use as a reinforcing agent for polymers in a number of applications; already commercial products have appeared in the market (e.g. graphene tennis racquets). For such applications it is vital to understand flake/ polymer interactions and the mechanisms through which stress is transferred from the surrounding polymer to the nano-inclusion. Bearing in mind the rather weak bonding between the two materials, another pertinent question refers to the level of applied stress for which interfacial debonding or sliding initiates.

In order to be able to assess the stress or strain built up along a graphene flake, it is necessary to employ a technique, such as Raman spectroscopy, that is capable to yield values of graphene strain through the shift of certain vibrational modes. In fact, previous work by us and others[11-13] has shown that the G -peak frequency ($\omega_G$) of monolayer graphene and carbon fibers relates linearly to tensile strain. Depending on graphene orientation, the imposition of a uniaxial stress field leads to the lifting of the $E_{2g}$ phonon degeneracy and the splitting of the $\omega_G$ peak to $\omega_G^-$ and $\omega_G^+$ components that exhibit relative large values of wavenumber shifts per strain of the order of $-31.4$ cm$^{-1}$/% and $-9.6$ cm$^{-1}$/%, respectively[11]. By comparing a whole range of graphitic materials such as carbon fibers and graphene, a universal stress sensor has been established[14] having wavenumber stress sensitivity of $-5\omega_G^{-1}$ (cm$^{-1}$MPa$^{-1}$), where $\omega_G$ is the peak position at zero stress for both graphene and carbon fiber with annular morphology. In addition to the above, the frequency position of the 2D Raman band ($\omega_{2D}$) shows even higher values of wavenumber shifts per strain of the order of $-60$ cm$^{-1}$/%[12, 13] depending on the excitation (laser) wavenumber which affects the value of $\omega_{2D}$ at rest.

Since, as already mentioned, the main Raman vibrational modes, $\omega_i$ *(i* stands for G, 2D*)* of graphene are related to applied strain then it is reasonably easy to use the obtained $\omega_i$ vs. strain relationship to resolve the reverse problem i.e. to convert Raman frequency values from a simply supported EG into a strain distribution along a given flake. The spatial resolution of such measurements is only limited by the wavelength of the incident laser beam and the optical arrangement used to focus the beam onto the specimen. Since significant strain fluctuations are expected to be encountered at distances less than the optical diffraction limit (~ 500 nm) then any scanning along the flake should be conducted at steps of the order of several tens of nanometers in order to capture the strain built up from any flake discontinuity such as a flake edge. This indeed represents a major challenge for conventional Raman measurements as the assumed sensitivity is considered to be approximately of the order of the diffraction limit. However, as it will be shown in this work notable differences are observed when a ~1 μm laser spot is translated by means of nanopositioning stages (see experimental section) at steps of 100 nm.

To date strain transfer measurements as a result of external mechanical loading have been conducted on embedded monolayer[15] and bi-layer graphene[16] on systems similar to those employed here. In all cases the flakes exhibited shear-lag type of stress built up upon the imposition of the external (tensile) stress field. However, due to the low resolution of the spatial Raman measurements, no safe conclusions could be drawn on the form of the strain built-up at the edges[16]. The first attempts to assess the stress transfer characteristics in exfoliated monolayer graphene deposited on an epoxy based photoresist (SU8)/poly(methyl methacrylate) (PMMA)/ substrate have been reported by Young et al.[15, 16] and on polyethylene terephthalate (PET) by Zhu et al.[17]. Both of them examined the stress transfer mechanism using the "*shear-lag*" principles proposed by Cox as early as 1952[18]. However, the spatial resolution of the measurements was a few microns (1-2 μm) in the best case and therefore there the fitting of the "*shear-lag*" curves through the data points appeared quite poor. Thus, any phenomena such as unintentional doping[19, 20] or even the stress transfer mechanism that take place within a distance of 1-2 μm from the edge of the flake could not be observed due to the relative poor spatial resolution of the measurements. Regarding the work reported by by Zhu et al. [17] on a PET substrate, it was postulated[17] that as the PET substrate is stretched, retraction of the matrix (sliding) led to failure, such as shear sliding under tension and graphene buckling under compression. However, the critical strain for which sliding occurs was not estimated directly as the authors assumed shear lag behavior in order to extract the above value from the center of the flake[17].

In this work, a detailed Raman mapping of a simply supported monolayer graphene (1LG) onto a SU8/PMMA matrix under incremental tension is carried out. Emphasis is given to the

stress/ strain characteristics near the edges of the specimen. It is shown for the first time that, the distribution of axial stress (strain) along the flake deviates somewhat from the classical shear-lag prediction for a region of 1–2 μm from the edge. This behavior is mainly attributed to the presence of residual stresses, unintentional doping and/or edge effects. The identification of phenomena that are prevalent at or near the edges of the flake stem from measurements conducted on 5 graphene/ SU8-PMMA systems. Overall, in these experiments we have observed the presence of (a) strong residual stresses due to the exfoliation process (see Supporting Information, fig. S2) and/or the morphology of the underlying substrate (b) unintentional doping at the edges (see Supporting Information, Table S1).

As stated in our early works [21], the strain transfer profiles obtained by the Raman technique can be converted into interfacial shear stress profiles along the length of the reinforcement by means of a straightforward balance of forces argument. This indeed captures the very essence of reinforcement in polymer composites that incorporate stiff inclusions (eg fibre, flake etc) since the prevailing mechanism is shear at the interface, which is converted into normal stress at the inclusion. Indeed, in the case of a flake, if we consider an infinitesimal flake length *dx* near its edge (see Supporting Information S5), then the stress equilibrium can be obtained. In the works of Young et al.[15, 16] and Zhu et al. [17]., the Cox's approximation[18] was employed since the data were not accurate enough to determine the stress derivative of equation (2) in Supporting Information S5. Indeed this is the main novelty of our approach: by conducting measurements at the nanoscale we can capture the built up of stress near the graphene edges and therefore we can obtain accurate estimates of the interfacial shear stress distribution along the flake.

## Results and Discussion

An optical microphotograph of the studied simply-supported graphene sample is presented in Fig.1. Initially, a detailed Raman mapping based on the 2D peak position of the pristine stress-free flake took place (Fig. 1b). As seen, within an area of the flake from -10 to -22 μm in the y- direction (Fig. 1b), the Pos(2D) along the strain axis is reasonably uniform with small fluctuations in small enclaves. In order to conduct detailed Raman measurements the edge-to-edge shift of the $\omega_{2D}$ peak along a line at approximately y = -20 μm of graphene flake (Fig. 1a) was monitored. The overall mapping length of ~10 μm was equal to the width of the flake at that position.

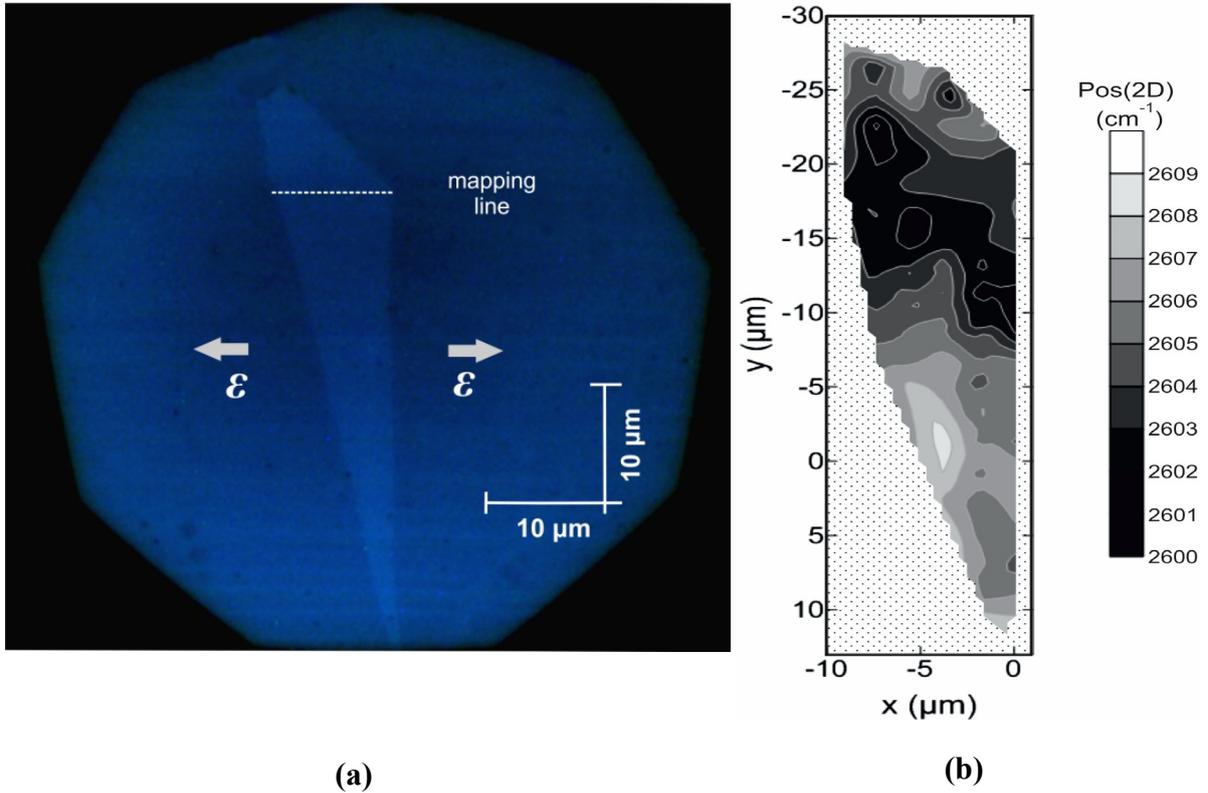

(a)          (b)

**Figure 1:** (a) Optical micrograph of the simply-supported monolayer graphene. The mapping line and the direction of strain are clearly marked at the trace in the middle. (b) A contour map of $\omega_{2D}$ peak for all the area of the examined simply-supported monolayer graphene

The experimental data of the frequency position for $\omega_G$ (Pos($\omega_G$)) and $\omega_{2D}$ mode (Pos($\omega_{2D}$)) as a function of the distance from the graphene edge are plotted in Fig. 2 for the as-received specimen but also at various increments of tensile strains from 0.0% up to 1.00 %. As it will be shown below, the distributions of Pos($\omega_G$) and Pos($\omega_{2D}$) along the mapping line for each incremental strain can be converted to strain distributions. Furthermore, from the ratio of their intensities information on the doping level can be retrieved.

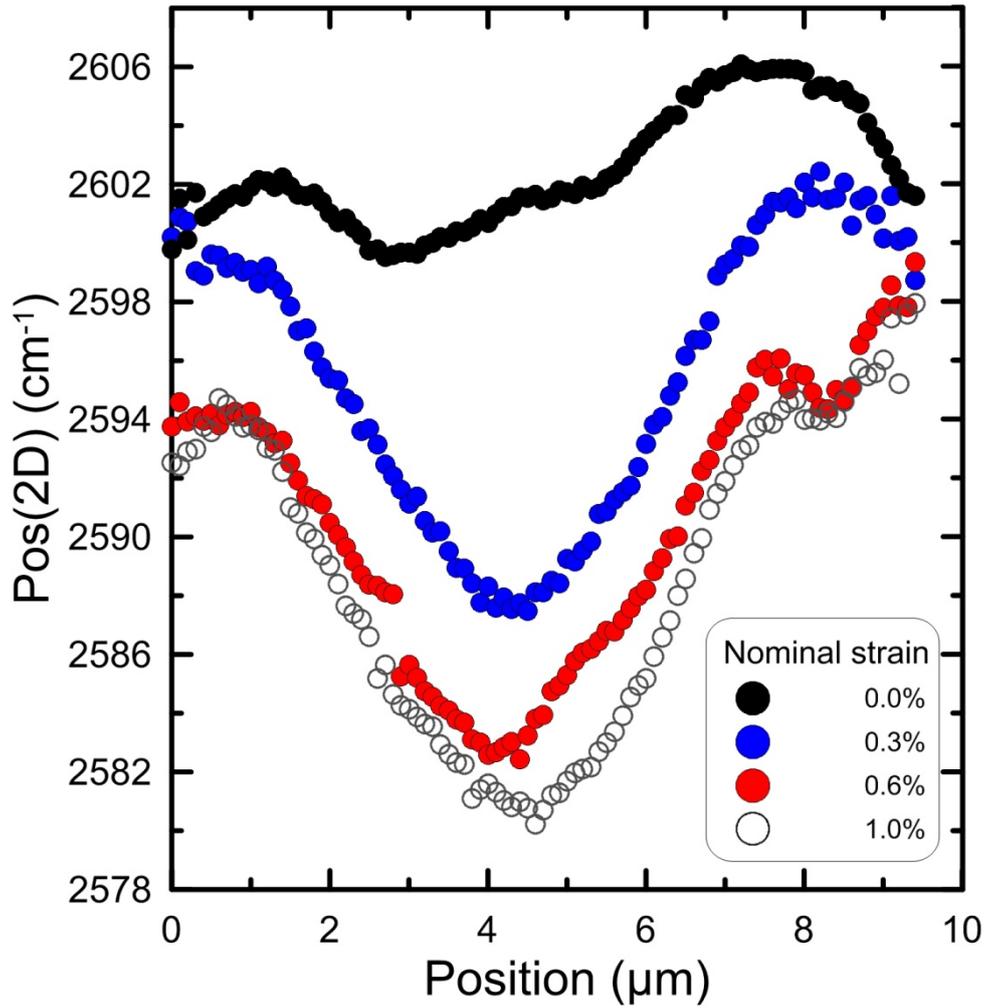

(a)

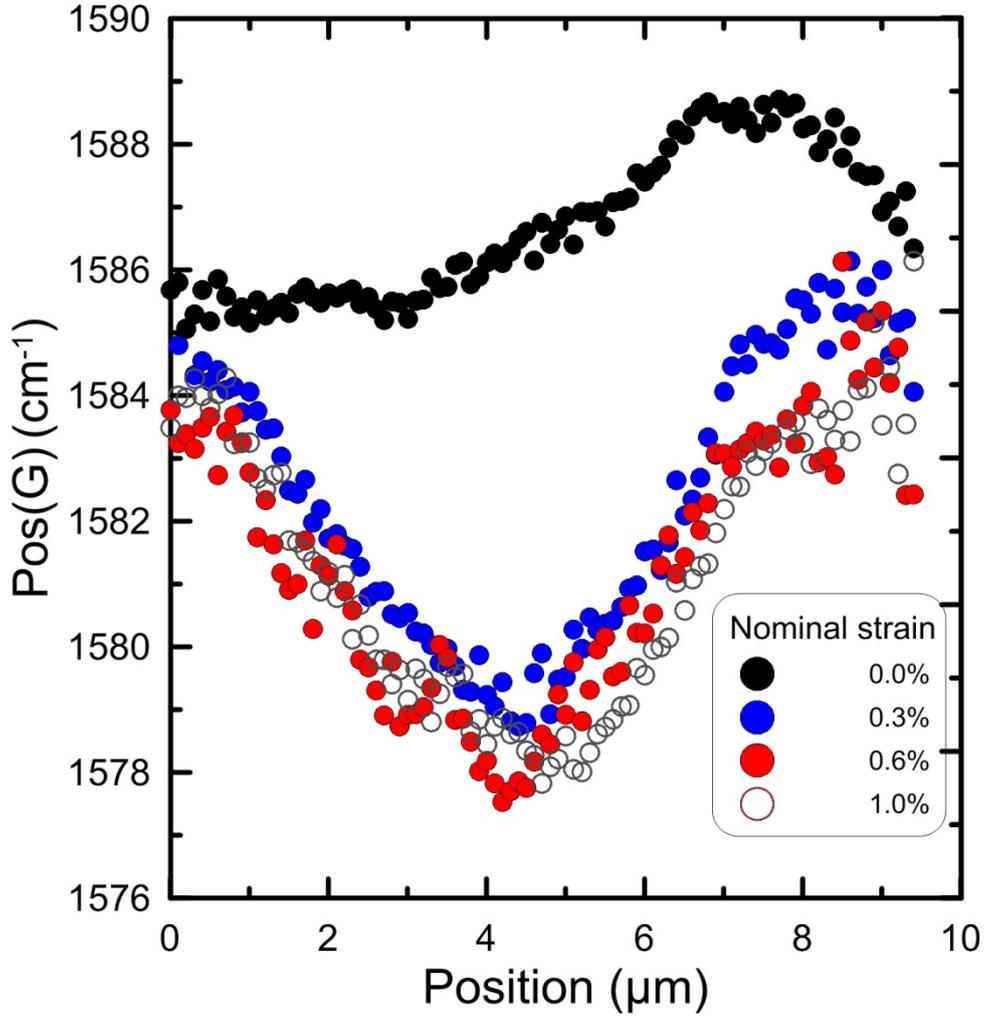

**(b)**

**Figure 2:** The **(a)** Pos($\omega_{2D}$) and **(b)** Pos($\omega_G$) distributions along the mapping line at various levels of strain

It is important to note that for a purely deformed graphene sheet (no significant doping effects), a linear relationship between Pos($\omega_{2D}$) vs. Pos($\omega_G$) is expected of slope of around 2.3 – 2.5[13, 22]. Now by correlating Pos($\omega_{2D}$) vs. Pos($\omega_G$) as shown in Fig. 3, it can be argued that most collected data points are under mechanical loading since the majority of the points follows a linear dependence. However, the least-squares-fitted slope of 1.9 is slightly lower than the expected range of 2.3-2.4 and this indicates that a small amount of unintentional doping is present. This emanates from the production procedure, as well as, from the interaction with the substrate and the environment (adsorbents, resist/process residuals, etc.) [22,23,24].

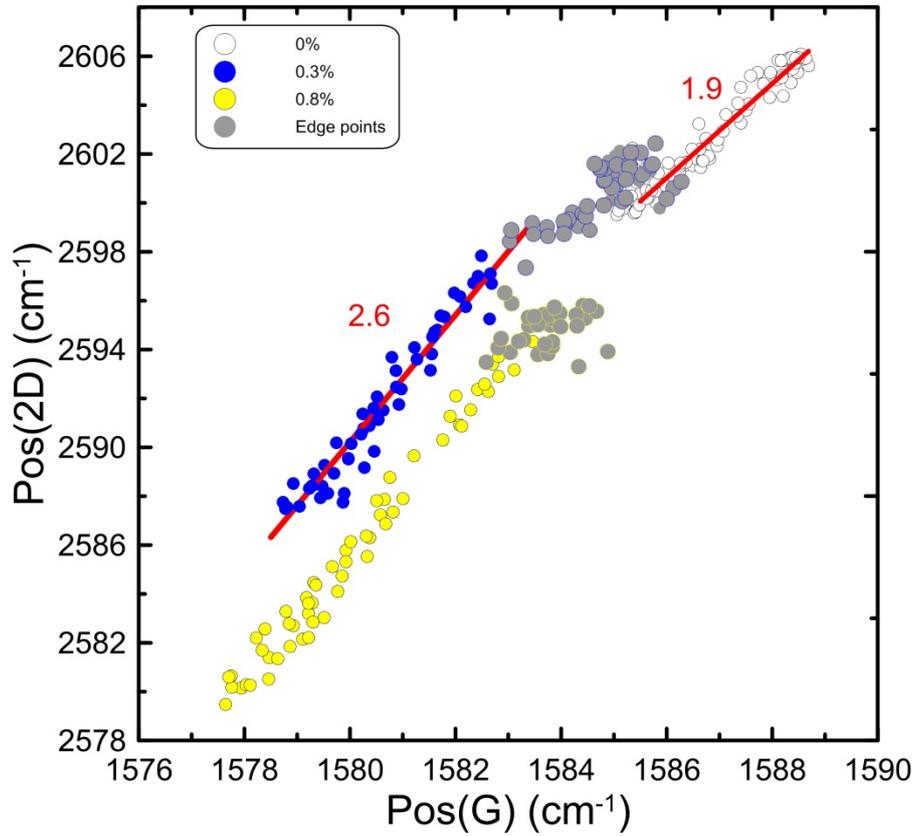

**Figure 3:** The correlation of Pos($\omega_{2D}$) and Pos($\omega_{2G}$) at various levels of applied strain. The grey points correspond to mapping points for distances from the edge of the flake of ~ 1.5 μm.

Das et al[25] have shown that for low doping levels, the Pos($\omega_{2D}$) does not change and can be used for sensing the uniaxial strain applied onto the graphene flake. Taking into account that the free standing graphene is almost charge neutral and stress free, the $\omega_G$ and $\omega_{2D}$ bands appear at ~1580 and ~ 2680 cm$^{-1}$, respectively for $E_{laser}$ = 2.41 eV (514 nm)[26]. By using the excitation energy dispersion factor of -100 cm$^{-1}$/eV, the $\omega_{2D}$ band for charge neutral and stress free graphene is expected at 2595 cm$^{-1}$ for $E_{laser}$ = 1.58 eV (785 nm) used in our experiments[27]. Therefore, the mapping points at 0.0% of tensile strain are under residual compressive strain, which varies from 0.1 % at the edges up to 0.2 % at positions within the flake (Fig.2). By subtracting the strain contribution from the Pos($\omega_G$) a mean value of about

±3x10$^{12}$ cm$^{-2}$ of unintentional excess charge was found which corresponds to a shift in Fermi energy from the neutrality point by about 180 meV[28].

For strain levels greater than 0.30%, and at locations between 3 and 5 μm, where strain reaches its maximum value, the data points for the Pos($\omega_G$) peak are obtained as a weighted average of the corresponding data of the two components, since $\omega_G$ band splits into two distinct components (G$^+$ and G$^-$) upon the application of uniaxial tensile strain[11, 12]. For points located at distances less than 3 μm and greater than 5 μm, no splitting is observed and the corresponding data were fitted with a single Lorentzian.

As seen, systematic shifts of Pos($\omega_{2D}$) peak are obtained as one moves in steps of 0.1 μm nm from the edge of the flake towards the middle. The observed shifts can be attributed to stress/strain effects and, in some cases, to doping which results from the presence of impurities at various positions along the mapping line (Fig. 1a). It has been found both theoretically[28] and experimentally[25] that the phonon frequency and lifetime of $\omega_G$ band exhibit a strong dependence on excess charge, due to the modification of phonon dispersion close to the Kohn anomaly. On the contrary[25], as mentioned above, $\omega_{2D}$ is almost unaffected for electron concentrations between ±0.5 x 10$^{13}$ cm$^{-2}$.

Casiraghi et al.[29], have shown that there is a remarkable stiffening of Pos($\omega_G$) as a function of the distance from the flake edges which are prone to doping[19]. The presence of doping at the edges can be easily identified in Pos(2D) *vs*. Pos(G) correlation graph (Fig. 3). For instance, at 0.0% of applied strain, a small amount of points (grey points) around Pos($\omega_G$) ~ 1585 cm$^{-1}$, which corresponds to a mapping length of 1.5 μm from the edge, does not follow the linear correlation. At 0.3% of applied strain, these points (blue-grey) exhibit a small red-shift having the same value of Pos($\omega_{2D}$) at ~ 2600 cm$^{-1}$. The same applies to the distribution of 0.8% strain as can be seen by the yellow-grey points, which are blue shifted by about 5 and 2.5 cm$^{-1}$ for Pos(2D) and Pos(G), respectively.

Since doping shifts the Fermi energy, the intensities of both $\omega_G$ and $\omega_{2D}$ bands are also altered[30]. Thus, the ratio of the frequency-integrated areas under $\omega_{2D}$ and $\omega_G$ peaks, A(2D)/A(G), can be used to differentiate doping within the mapping area since the lowest values of the ratio correspond to higher doping levels[31]. It is found that for the strain levels mentioned as above, the A(2D)/A(G) ratio of the grey points is lower by about 30% relative to other mapping points which follow the linear correlation of the Pos(2D) vs Pos(G).

The Pos($\omega_{2D}$) and Pos($\omega_G$) versus the applied axial strain are plotted in Figs 4a and b respectively. Each data point results from the averaging of the corresponding values within a range of 1.5 μm (similar to the spot size) centered at the peak of the Pos($\omega_{2D}$) and Pos($\omega_G$) distributions located at a distance of 4.5 μm (Fig. 3). Both Pos($\omega_{2D}$) and Pos($\omega_G$) red-shift linearly at a rate of -54.6 and -20.2 cm$^{-1}$/%, respectively, for strains up to 0.4% while at higher deformations (up to 1%) they reach a plateau. It is thus obvious that for strains up to 0.4 % the substrate/ graphene adhesion permits sufficient strain transfer by shear. For higher strains up to 1%, the substrate/ graphene interface fails and the flake slips. The latter was confirmed by unloading the specimen and mapping again the examined line of Fig. 1a (see Supporting Information, fig S3).

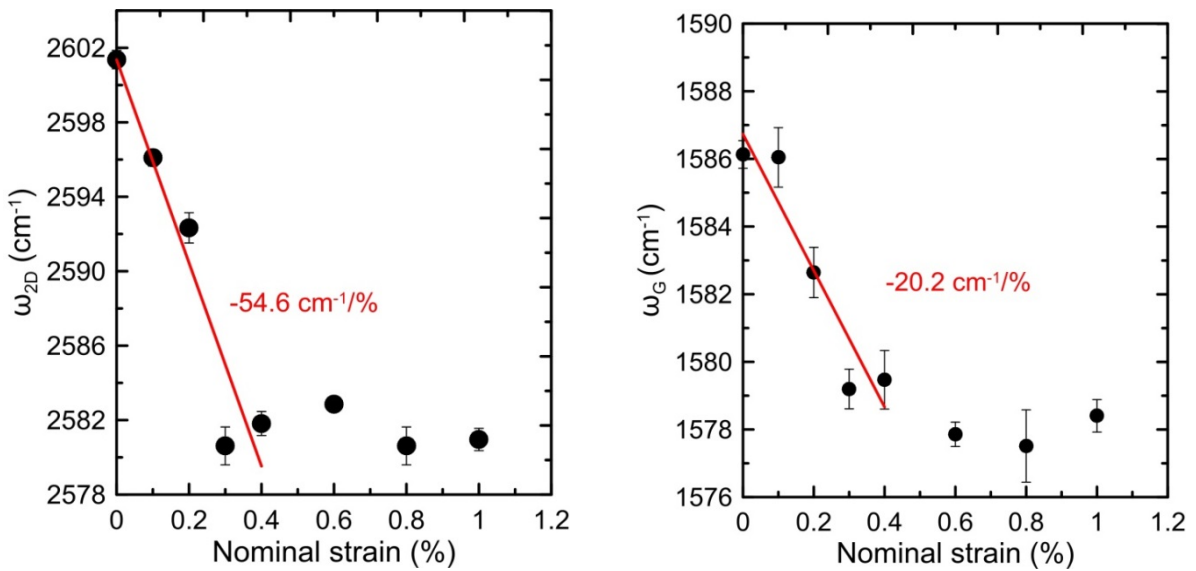

**Figure 4**: The (a) Pos($\omega_{2D}$) and (b) Pos($\omega_G$) peak positions vs. the applied strain

Regarding the Pos($\omega_{2D}$) and Pos($\omega_{2G}$) as a function of distance from the edge (Fig. 2), there is a gradual change of the sign of the slope from positive to negative over a distance of 1.5 μm from both edges indicating that most of the compression is gradually relaxed and the flake in that region is eventually subjected to tension. At higher strain levels, it is quite evident that the region from 2 to 4.5 μm (Fig. 2) appears to be free of residual strain and shows the highest rate of tensile stress take up. The region on the right hand side of the flake (Fig. 2) is already in compression and therefore lags behind the rest of the flake. An example of a graphene flake with similar sensitivities for the two phonons ($\omega_G$ and $\omega_{2D}$) is presented in the Supporting Information (Simply supported case (flake 2)).

To estimate the real strain applied to the graphene flake, the exact Pos(2D) peak value of the graphene at 0.0% of applied strain should be known. In this case, we consider the weighted average of Pos(2D) values of all the data points along the mapping line that were located within the region 3 to 5 μm as a representative $\omega_{2D}$ value in the absence of external loading. This is found to be 2602.4 ±1.9 cm$^{-1}$. Then using the Pos(2D) at each mapping point, the corresponding strain, $\varepsilon$, was estimated from the following relationship:

$$\varepsilon = \frac{(\omega_{2D}|_\varepsilon - \omega_{2D}|_{\varepsilon=0.0\%})}{k_{2D}} \quad (1)$$

where $\omega_{2D}|_{\varepsilon=0.0\%}$ = 2602.4 ±1.9 cm$^{-1}$, $\omega_{2D}|_\varepsilon$ is the corresponding strain, $\varepsilon$, at each measured point and $k_{2D}$ = -61.9 cm$^{-1}$/%[32], the dependence of $\omega_{2D}$ Raman phonon shift with the strain

In figure 5, the strain distributions for various levels of applied tensile strain derived from the shifts of the Raman wavenumber, are shown. The edges of each side of the flake are taken as the zero distance points, in order to depict the evolution of flake's strain with the applied strain. It should be noted that the results presented below are independent from the choice of the absolute value of the Pos(2D) at 0%.

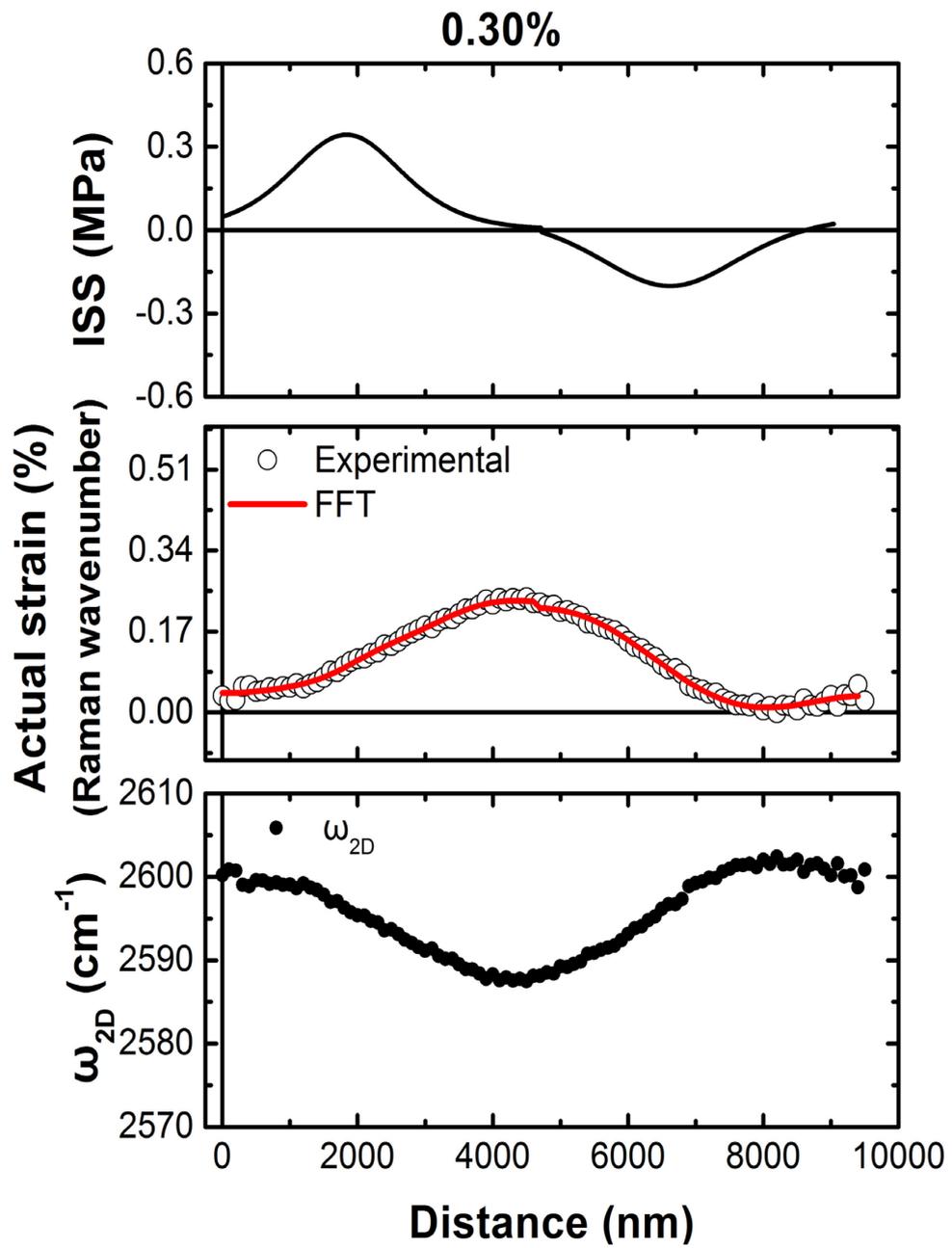

(a)

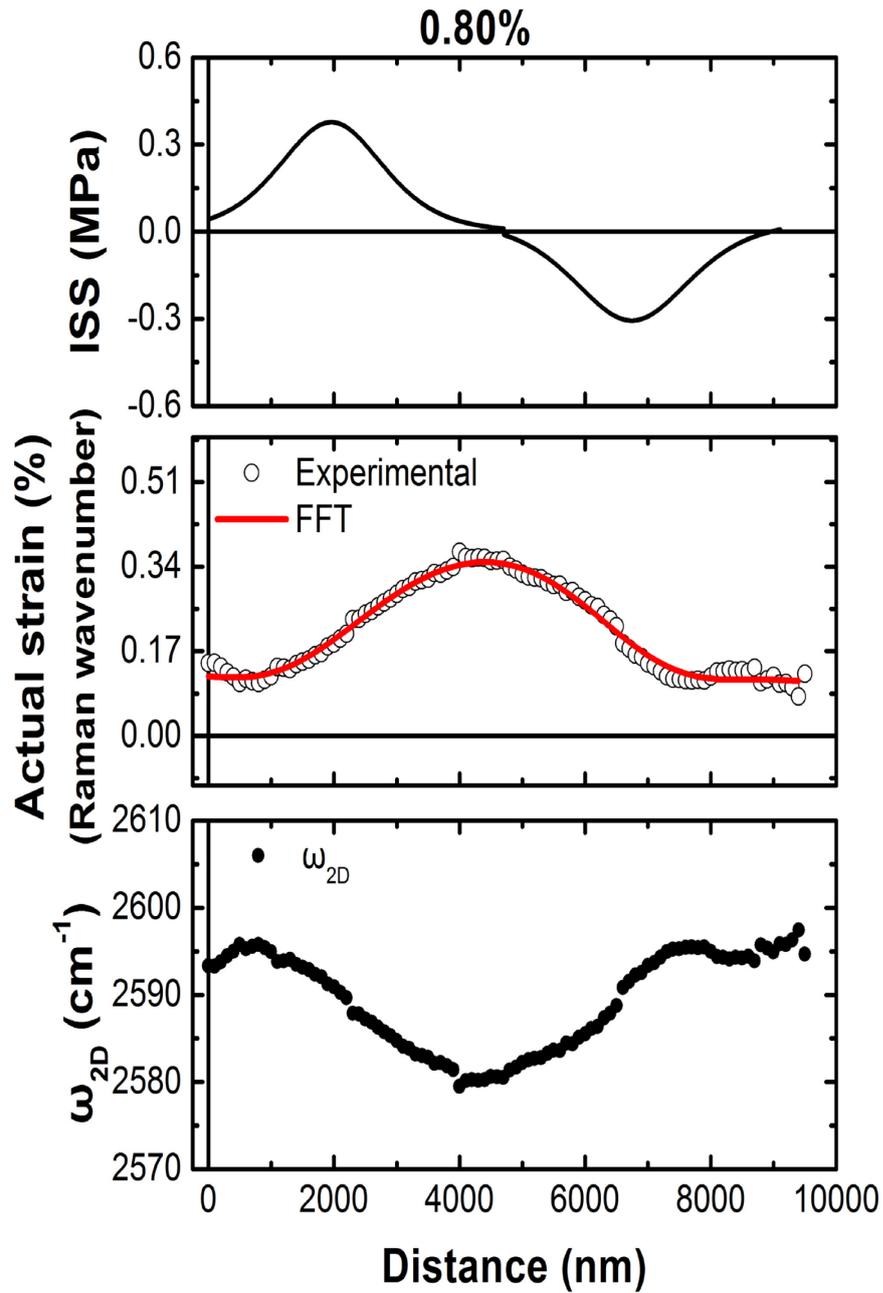

(b)

**Figure 5:** <u>Bottom</u>: Raman wavenumber distributions of the $\omega_{2D}$ peak for the simply supported case at applied strains of (a) 0.30% and (b) 0.80%. <u>Middle</u>: The resulting axial strain distributions via the Raman wavenumber shift for (a) 0.30% and (b) 0.80%. The red solid line is a guide to the eye. <u>Top</u>: The corresponding interfacial shear stress distributions along the whole length of the flake for (a) 0.30% and (b) 0.80%.

Similar strain distributions are extracted using as a calibration factor the average of $\omega_G^-$ and $\omega_G^+$ components[11] for strains up to 0.30% (-19.4 cm$^{-1}$/%) (see supporting information, Fig S4). Uniaxial stress in the substrate leads to the development of interfacial shear between graphene and polymer, which, in turn, is converted into a normal stress in the flake. This force transmitted though shear across the flake/ substrate interface is acting primarily near the edges of the flake. By simply balancing the shear-to- axial forces[33, 34] (see also Supporting Information, fig S5), we obtain:

$$\frac{d\sigma}{dx} = -\frac{\tau_t}{nt_g} \quad \text{or} \quad \frac{d\varepsilon}{dx} = -\frac{\tau_t}{nt_g E} \quad (2)$$

where $\sigma$ is the axial stress acting on the flake, $\tau_t$ is the interfacial shear stress between graphene and polymer, $n$ is the number of graphene layers ($n$=1), $E$ is the modulus of graphene (1 TPa)[35] and $t_g$ is the thickness of the monolayer[36-38].

The Interfacial Shear Stress (ISS) distributions in the flake at strains of 0.3% and 0.80 % are presented in Figs. 5 and S4. At this point, it should be stressed that for distances ~1500 nm from the flake edges, the obtained ISS profiles should be considered with caution since the mechanical effect is affected by the presence residual strains and doping. For all strain levels, ISS reaches its maximum at a distance of about 2 μm from the flake edges (Fig. 5). Due to the short width of the flake over which mapping took place, the ISS reaches zero values only within a very short length in the middle of the flake. Moreover, the maximum ISS of graphene/ polymer interface is not achieved at the edges of the flake as expected for simply supported single layer graphene under strain[17] but at a distance of about 2 μm from the edges. A comparison between the expected classical Cox type strain distribution[18] and the corresponding strain distribution of the present study is made in Figure S6. Since the strain rises to about 90% of the plateau value (0.30% applied strain) over about 1.5 μm from the edge of the flake[16], it seems that the classical shear-lag model[18] cannot describe properly the experimental stress distribution. As mentioned above, a part of the flake in the vicinity of the edges has been unintentionally doped, while a significant amount of residual compressive strain developed during the production procedure. Both effects are clearly detected in the Pos(G) vs. Pos(2D) correlation diagram shown in Fig.3, in which the grey circles correspond to points with similar Pos(2D) values within a range of 3-4 cm$^{-1}$ for Pos(G). Therefore, they are not linearly correlated (slope ~ 2.5) unlike the majority of the data points in the mapping region which are under tensile strain. It is, thus, concluded that both the presence of residual stresses (compressive) and the unintentional doping at the flake edges affect the level of graphene/ polymer adhesion which in turn alters the stress/ strain transfer characteristics.

In Fig. 6, the measured ISS$_{max}$ at a distance of about 2 μm from the flake edges as a function of the tensile strain is presented. Both the Pos(G) and Pos(2D) distributions at various strain levels were used for the evaluation of ISS$_{max}$ using equation 2.

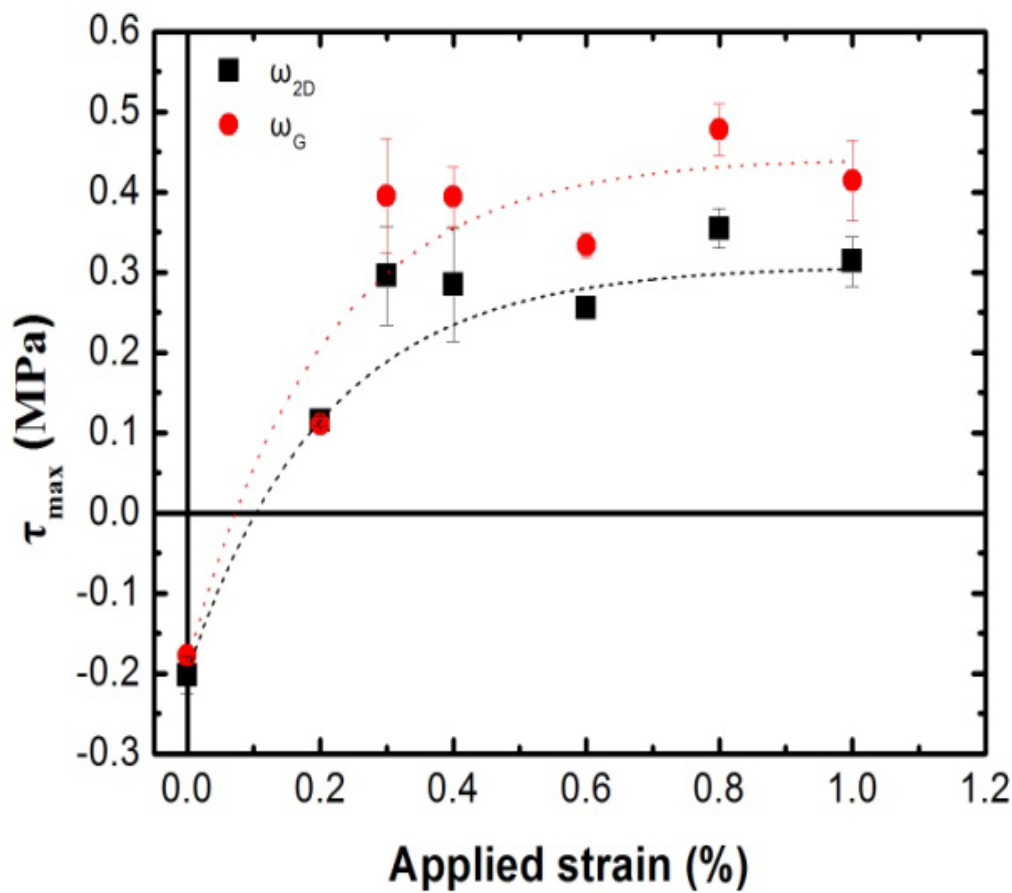

**Figure 6**: Maximum Interfacial Shear Stress (average values for both edges of the examined flake) as a function of the applied tensile strain extracted from $\omega_G$ and $\omega_{2D}$ bands. The dot lines are a guide to the eye

As can been seen $ISS_{max}$ reaches a plateau of ~0.5 MPa at a strain of 0.4%. Any differences between the values of both bands should be attributed to the effect of doping upon the $\omega_G$. It is worth mentioning that the interfacial shear strength values obtained here compare reasonably well with those obtained by Zhu et al.[17] on PET substrates. This is not surprising since in all such systems[11-14, 39-47], the interfacial stress transfer takes place primarily through weak van der Waals interactions and therefore the effect of polymer chemical structure is minimal.

Additionally, the derived ISS distribution differs significantly from the typical profiles that have been presented in fibre/polymer systems[48]. As seen in figs. 2, 5, and S4, there is a hysteresis in the strain profiles up to distance of 1.5 μm from the edges. Within this area, the corresponding response of $\omega_{2D}$ vs. $\omega_G$ is not linear (points with similar $\omega_{2D}$ values within a range of 3-4 cm$^{-1}$ for $\omega_G$) (fig. 3). As stated elsewhere[19], doping effects can cause such behavior. Therefore, the edges, and an area of micrometer dimensions away from the edges are clearly affected by non-mechanical interactions. Thus, the graphene/polymer adhesion (taking place through van der Waals bonding across the atomically thin surface) is weakened by the doping and this is clearly manifested by the low rate of stress transfer near the edges and the obtained low values of interfacial shear stress within the affected area. Provided that the inclusion is long enough (larger than two times the transfer length for stress transfer) then efficient reinforcement occurs[33, 34]. In the present experiment, however, the transfer length is the sum of the affected area due to the residual stress plus the length required for elastic stress transfer which amounts to values of 4 μm (left-hand) and 5 μm (right-hand) from the edges (Supporting Information, S9: Elastic stress transfer).

According to past work[49], for discontinuous model composites, the transfer length is defined as the distance from the edge where the interfacial shear stress tends to zero, provided that the reinforcement is long enough (at least 10 times greater than the transfer length) [16]. In the present work, the transfer length seems to be the sum of the edge area, where doping and/or edge effects are taking place, plus the required length for elastic stress transfer. Since, the total measured length was approximately 10 μm and the total transfer length was 9 μm, it is obvious that for efficient stress transfer, flakes much greater than ~10 μm are required. This is a very important conclusion for practical applications and may be valued not only for graphene/ PMMA composites but also for other polymer matrices for which inclusions of sizes greater than 10 μm are required for efficient reinforcement.

**Conclusions**

This study has shown by means of measurements conducted at the submicron level that the distribution of axial strain (stress) along a simply supported flake is affected by different phenomena that appear at the vicinity of flake edges. The presence of compression fields due to the preparation procedure (e.g. exfoliation) and doping species due to interaction of the flake with the substrate, can affect the stress transfer mechanism at an approximate distance of 2 μm away from the edges. As a result both the corresponding axial stress (strain) and interfacial shear stress distributions along the flake deviate from the expected classical shear-lag prediction. The transfer length for stress transfer from each side of graphene flake edge consists of a region affected by doping effects for which the stress transfer is poor and another region that is dominated by elastic, shear-lag type, effects. The overall length for efficient load transfer is estimated to be ~ 4 μm from each end of the flake which clearly means that for the simply supported case only flakes greater than 8 μm will be fully loaded. For large flakes a maximum value of interfacial shear stress (ISS) of 0.4 MPa is obtained prior to flake slipping. This value represents the upper limit of ISS that the graphene/ polymer system can endure. The results from fully embedded graphene/ polymer systems will be presented in a future publication.

# Experimental Method

The simply-supported monolayer graphene was prepared by mechanical cleavage from HOPG (High Order Pyrolitic Graphite) and transferred onto a cantilever polymer beam. Before the deposition of the flake on the poly(methyl methacrylate) PMMA beam, a thin layer of an epoxy based photoresist SU-8 2000.5 (of thickness $t_{SU8}$~200 nm) was spin coated on the top in order to increase the optical contrast of the flake. The thickness of the PMMA substrate was $t = 3.0$ mm and the investigated flakes were located at a distance, $L/2$, where L=80 mm is the length of the supporting span of a four-point bending apparatus frame (Fig.7).

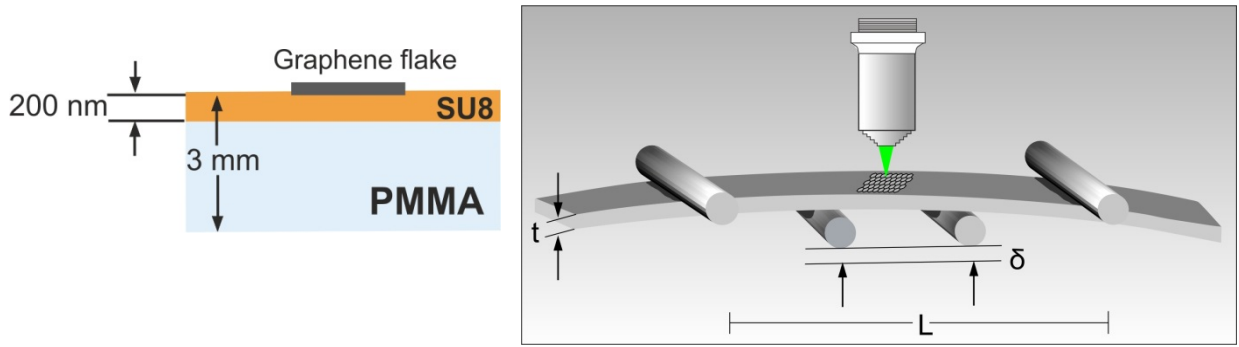

**Figure 7**: Schematic representation of the cross section area of the specimen used (left) and the 4-point bending apparatus (right)

The top surface of each beam was subjected to tension by flexing upwards the beam by means of an adjustable screw. The deflection $\delta$ was achieved by rotating a small lever connected to a set of gears and a long screw attached to the moving jaw of the 4-point bending apparatus. Slow travelling of the lever results in accurate incremental steps of applied strains to the specimens. In fact, it has been calculated that 10/8 of a full turn corresponds to an approximate elongation of ~0.45 mm, which corresponds to 0.10 % of applied strain according to the following equation.

$$\varepsilon(\delta) = 4.47 \frac{\delta t}{L^2} \quad (3)$$

where $\delta$ is the deflection (manually applied) on the PMMA bar, $L$ the length of supporting span and $t$ the thickness of the PMMA bar.

The validity of eq. 3 has been confirmed by independent measurements of axial strain obtained by attaching minute strain gauges within the middle area of pure PMMA beams. Finally, since the relationship between the $\omega_{2D}$ peak and strain has already been established earlier[11, 13, 32] the external strain can also be estimated from the shift of the $\omega_{2D}$ peak at each increment of strain with reference to the position of the peak at 0% strain. The four-point bending apparatus was placed on a 3-axis piezoelectric translation stage and a piezoelectric controller from Thorlabs Inc. The NanoMax 3 axis flexure stage can provide nanometric positioning on three orthogonal axes[50].

At each strain level the stage was moved with a step of 0.1 μm (simply supported) collecting simultaneously Raman spectra, thus a performing a detailed mapping across a specific line on the examined flakes. The line mapping strategy followed here has been used successfully in the study of local mechanical stress measurements by micro Raman spectroscopy in silicon devices[51, 52]. Raman spectras are measured at 785 nm (1.58eV) for the single supported case, using a MicroRaman (InVia Reflex, Rensihaw, UK) set-up. The laser power was kept below 1.0 mW on the sample to avoid laser-induced local heating, while an Olympus MPLN100x objective (NA = 0.90) was used to focus the beam on the samples.

By moving the laser spot at increments of 100 nm starting from graphene's edge and moving inwards as shown in fig.S1, we were able to capture the shift of the $\omega_{2D}$ peak as a result of the uniaxial (strain built up along that region which is not steep (see Fig.1). The sharp interface between graphene and PMMA allow us to estimate the actual laser spot size by recording series of Raman spectra across the interface (fig S1).

## ASSOCIATED CONTENT

## Supporting Information

Further experimental data and explanations are given upon the estimation of the actual laser spot size, the unintentional doping at the area close to the edges and the indications of interface failure. Also, the implementation of balance of forces on single monolayer graphene and the differences from the expected classical shear-lag distribution is shown. This material is available free of charge via the Internet at http:// pubs.acs.org.

## AUTHOR INFORMATION

### Corresponding Author

*Whom all correspondence should be sent to: c.galiotis@iceht.forth.gr, kpapag@iceht.forth.gr

### Author Contributions

The manuscript was written through contributions of all authors. All authors have given approval to the final version of the manuscript.

## ACKNOWNLEDGEMENTS

The authors would like to acknowledge the financial support of Graphene FET Flagship (''Graphene-Based Revolutions in ICT and Beyond''- Grant agreement no: 604391) and the "Tailor Graphene" ERC Advanced Grant (no: 321124).

Dr. Polyzos would like also to acknowledge the support of the action "Supporting Postdoctoral Researchers" of the Operational Program "Education and Lifelong Learning" which is co-financed by the European Social Fund (ESF) and the Greek State through the General Secretariat for Research and Technology.

48. Anagnostopoulos, G.; Parthenios, J.; Andreopoulos, A. G.; Galiotis, C. An Experimental and Theoretical Study of the Stress Transfer Problem in Fibrous Composites. *Acta Materialia* 2005, 53, 4173-4183.

49. Jahankhani, H.; Galiotis, C. Interfacial Shear Stress Distribution in Model Composites, Part 1: A Kevlar 49® Fibre in an Epoxy Matrix. *Journal of Composite Materials* 1991, 25, 609-631.

50. Labs, T. Max 300 Series Nanomax 3-Axis Flexure Stage, User Guide. Labs, T., Ed.

51. De Wolf, I.; Maes, H. E.; Jones, S. K. Stress Measurements in Silicon Devices through Raman Spectroscopy: Bridging the Gap between Theory and Experiment. *J Appl Phys* 1996, 79, 7148-7156.

52. Wolf, I. D. Micro-Raman Spectroscopy to Study Local Mechanical Stress in Silicon Integrated Circuits. *Semiconductor Science and Technology* 1996, 11, 139.
ACS AuthorChoice - This is an open access article published under an ACS AuthorChoice License, which permits copying and redistribution of the article or any adaptations for non-commercial purposes.George Anagnostopoulos, Charalampos Androulidakis, Emmanuel N. Koukaras, Georgia Tsoukleri, Ioannis Polyzos, John Parthenios, Konstantinos Papagelis, Costas Galiotis, *Stress Transfer Mechanisms at the Submicron Level for Graphene/Polymer Systems*, ACS Applied Materials and Interfaces 7 (2015) 4216-4223, doi: 10.1021/am508482n30